\documentclass{article}

\usepackage{PRIMEarxiv}
\usepackage{makecell}
\usepackage[utf8]{inputenc} % allow utf-8 input
\usepackage[T1]{fontenc}    % use 8-bit T1 fonts
\usepackage{hyperref}       % hyperlinks
\usepackage{url}            % simple URL typesetting
\usepackage{booktabs}       % professional-quality tables
\usepackage{amsfonts}       % blackboard math symbols
\usepackage{nicefrac}       % compact symbols for 1/2, etc.
\usepackage{microtype}      % microtypography
\usepackage{lipsum}
\usepackage{fancyhdr}       % header
\usepackage{graphicx}       % graphics
\graphicspath{{media/}}     % organize your images and other figures under media/ folder
\usepackage[table,xcdraw]{xcolor}
\newcolumntype{L}[1]{>{\raggedright\arraybackslash}m{#1}}

\title{A Survey of Accessible Explainable Artificial Intelligence Research}

\author{
    Chukwunonso Henry Nwokoye\textsuperscript{1,2},
    Maria J. P. Peixoto\textsuperscript{1,2},
    Akriti Pandey\textsuperscript{3},
    Lauren Pardy\textsuperscript{3},
    Mahadeo Sukhai\textsuperscript{3},\\
    \textbf{Peter R. Lewis\textsuperscript{2}} \\
    \textsuperscript{1}Both individuals have equally contributed as the primary authors. \\
    \textsuperscript{2}Dept. of Business and Information Technology, Ontario Tech University, Oshawa, Ontario, Canada. \\
    \textsuperscript{3}Canadian National Institute for the Blind (CNIB), Toronto, Ontario, Canada.
}

\begin{document}
\maketitle

\begin{abstract}
The increasing integration of Artificial Intelligence (AI) into everyday life makes it essential to explain AI-based decision-making in a way that is understandable to all users, including those with disabilities. Accessible explanations are crucial as accessibility in technology promotes digital inclusion and allows everyone, regardless of their physical, sensory, or cognitive abilities, to use these technologies effectively. This paper presents a systematic literature review of the research on the accessibility of Explainable Artificial Intelligence (XAI), specifically considering persons with sight loss. Our methodology includes searching several academic databases with search terms to capture intersections between XAI and accessibility. The results of this survey highlight the lack of research on Accessible XAI (AXAI) and stress the importance of including the disability community in XAI development to promote digital inclusion and accessibility and remove barriers. Most XAI techniques rely on visual explanations, such as heatmaps or graphs, which are not accessible to persons who are blind or have low vision. Therefore, it is necessary to develop explanation methods through non-visual modalities, such as auditory and tactile feedback, visual modalities accessible to persons with low vision, and personalized solutions that meet the needs of individuals, including those with multiple disabilities. We further emphasize the importance of integrating universal design principles into AI development practices to ensure that AI technologies are usable by everyone.
\end{abstract}

% keywords can be removed

\section{Introduction and Motivation}
Explainable Artificial Intelligence (XAI) refers to the development of AI systems that offer clear and easy-to-understand explanations for the decisions or actions they take \cite{Davidleslie2020}. With AI's ever-increasing importance in our daily lives, it is crucial to ensure that these explanation technologies are accessible to everyone, including those with sight loss and other disabilities, to promote digital inclusion. We argue that while the field of XAI is rapidly growing, there is a lack of focus on accessibility in its design. To address this, we  propose the notion of \textit{\textbf{Accessible Explainable Artificial Intelligence (AXAI)}} in this paper. Achieving AXAI will remove barriers and ensure that clear and understandable explanations of AI actions and decisions can help people effectively use these technologies in their daily lives. Emphasizing the importance of AXAI can encourage the adoption of universal design practices, ensuring that AI technologies are designed to be used by everyone, regardless of their physical, sensory or cognitive abilities.

Besides ethical principles \cite{Nikolinakos2023}, the essentiality of explanations for attaining transparency, trust, reliability or dependability for AI systems has been strongly emphasized in the European Union’s Ethics Guidelines for Trustworthy AI \cite{ethics_guidelines}. In fact, research evidence has shown that counterfactual explanations can assist comprehension and contesting of AI decisions by citizens \cite{Laux2024}.  Also, Bahalul et al. \cite{Balul_HAQUE2023122120} highlighted the “\textit{right to explanation}” (Goodman and Flaxman, \cite{GoodmanBryce}) alongside AI system’s explainability as underscored by the General Data Protection Regulation. In the context of XAI, this “\textit{right to explanation}” is yet to be sufficiently defined and discussed for disability (visual, speech, etc.). Considering Ali, et  al., \cite{ALI2023101805}, explainability gives a comprehensible explanation of the model's choice, thus expressing what is happening inside the model.  Additionally, explanation implies ``an understanding of how a model works, as opposed to an explanation of how the world works'' \cite{Rudin2019}. Nevertheless, it is challenging to define exactly what constitutes an explanation. There have been several successful attempts at defining explanations in XAI research~\cite{Balul_HAQUE2023122120}, \cite{ALI2023101805}, \cite{Das2020}. For instance, Bahalul et al. \cite{Balul_HAQUE2023122120} presented the accepted meanings and explanations for several domains of research, but the research unfortunately lacks an accessibility and disability perspective. The affected domains include e-commerce, human resource management, digital assistants, e-governance, healthcare, media and entertainment, education, transportation, finance, and social networking. 

Going by the meanings and styles of the domains mentioned above, it is necessary to define and discuss explanations and XAI presentations from a disability point of view. This is premised on the positions of Ali et al. \cite{ALI2023101805}, which asserted that several academic research groups have studied the idea of explainability in decision-making by AI. However, every research community takes a different perspective on the problem and offers interpretations with distinct connotations. In this discourse, two significant communities seem to emerge, the technology and the disability community. By the \textbf{\textit{technology community}}, we are referring to groups, meetings, conferences, researches, developments, initiatives and endeavours, whether academic or industry, aimed at promoting AI, XAI, machine learning (ML), deep learning (DL). Additionally, the \textit{\textbf{disability community}} focuses on groups, meetings, conferences, researches, developments, initiatives and endeavours aimed at making technology accessible for individuals living with any form of disability. By implication, the \textbf{\textit{technical aspects}} of XAI are the focus of the technology community, while \textbf{\textit{human-centred aspects towards accessibility}} are the focus of the disability community. Through extensive XAI research, the technology community has made immense progress in defining and implementing explanations for various domains \cite{Balul_HAQUE2023122120}, \cite{ALI2023101805}. So, our focus is on the disability community. 

Therefore, to ensure the inclusion of the disability community in the XAI discourse, this paper also aims to connect findings to XAI and tools for accessibility. We also sought promising applications/adaptations from highly related studies in human-centred computing and accessibility, for instance, some identified papers for visual disability. When we say visual disability, it is crucial to understand that we mean persons who are \textbf{\textit{blind, Deaf-Blind, with low vision, as well as individuals with visual processing disabilities}}. To achieve this, we conducted a literature survey to evaluate how the accessibility of AI explanations, especially for people with sight loss, has been addressed.

\section{A Literature Survey of AXAI}
Many XAI techniques rely on visual explanations, such as heatmaps or graphs \cite{RudreshDwivedi2023}, which are not accessible to people who are blind or have low vision. As a result, making complex AI explanations understandable and accessible through non-visual modalities, such as audio or haptic feedback, is a significant challenge. Furthermore, the accessibility needs can differ significantly among people with visual impairments, necessitating personalized solutions. After reviewing academic literature, we present our methodology, results, analysis, and discussions related to AXAI.

\subsection{Methodological Approach}
We conducted a search for relevant literature on accessibility and explainable artificial intelligence using reputable databases. For this purpose, we have selected databases based on the quality and scope of the papers indexed by each. This is particularly important for interdisciplinary fields like AI and accessibility. The following databases have been chosen:

\begin{itemize}
    \item \textbf{Scopus} is known for its broad interdisciplinary coverage, including a range of high-quality journals in science, technology, medicine, arts and humanities.
    \item \textbf{ISI Web of Science} has a collection of high-impact publications, which is essential for capturing cutting-edge research in all areas of science and technology, including studies on explainable AI and accessibility.
    \item \textbf{IEEE Digital Library} provides access to high-quality technical literature on AI, machine learning, and their applications in accessibility and inclusion, specializing in engineering and technology.
    \item \textbf{Google Scholar} aims to include relevant works beyond traditional journals, covering a wide range of academic sources of diverse levels of quality.
    \item \textbf{El Compendex} is a comprehensive engineering database that ensures technical literature coverage in specific engineering areas related to AI and accessibility.
    \item \textbf{ACM Digital Library} provides access to research in assistive technology and explainable AI, as it is focused on computing and information technology literature.
    \item \textbf{CINAHL \& PubMed} offer access to an extensive collection of health-related literature.
\end{itemize}

We created search strings for our selected databases to capture the intersection between Explainable AI (XAI) and Accessibility, ensuring the inclusion of relevant work for people with different visual disabilities. Our search strings include key terms and synonyms, such as accessibility, disability, visual impaired, sight loss, low vision, blind, and deaf blind, to cover the range of terminologies used in the literature.

Search strings:
\begin{itemize}
    \item (``accessib*" OR ``inclusiv*") AND (``explainabl*" OR ``XAI" OR ``transparent" OR ``interpretab*") AND (``artificial intelligence" OR ``machine learning" OR ``deep learning") AND (``explanations" OR ``narration").
    \item (``blind" OR ``low vision" OR ``deaf blind" OR ``sight loss" OR ``vision loss") AND (``explainabl*" OR ``XAI" OR ``transparent") AND (``artificial intelligence" OR ``machine learning" OR ``deep learning").
    \item (``disabil*" OR ``visual impairment") AND (``explainabl*" OR ``XAI" OR ``transparent") AND (``artificial intelligence" OR ``machine learning" OR ``deep learning").
\end{itemize}

We conducted a systematic review using the Parsifal \footnote{Parsifal: \url{https://parsif.al/}} tool to manage papers and exclude duplicates from multiple databases. After using the specific search strings above and excluding duplicates, we initially selected 727 papers from 2019 to 2024. We then selected eight articles based on their title and abstract and analyzed them individually, leading to only four relevant papers. 

While the search string resulted in a large number of papers, many were in fact not related to AI technologies that are accessible and explainable for individuals with sight loss. We therefore used an exclusion criteria to remove works that do not address the accessibility of AI explanations, or that develop AI-based technologies for accessibility but without considering the importance of explainability.

As an example of the removed papers based on our exclusion criteria, consider the work titled ``Artificial Intelligence, Accessible and Assistive Technologies''~\cite{Draffan2020}. That paper discussed the potential of artificial intelligence (AI) to enhance accessibility and assistive technologies for individuals with disabilities. However, it fails to address transparency and explanation of AI decision-making. As a result, we did not incorporate that paper and similar ones in our research.

The exclusion also occurred for research papers that utilize AI-based technologies to create activities or products that are only accessible to particular groups of people rather than those with sight loss. For example, the study ``AI Enabled Tutor for Accessible Training" \cite{Banerjee2020}, which developed an AI tool to offer job training to individuals with disabilities, specifically the deaf and hard of hearing community, and similar research works were disregarded.

Other examples of exclusion can be seen in the use of the term ``accessible" to refer to a product or service that is affordable, as in ``Accessible Video Analytics: the Use Case of Basketball" \cite{Natale2022}, and something easy to use for the general public, but not explicitly designed for sight loss people, as in ``Building XAI-Based Agents for IoT Systems" \cite{Dobrovolskis2023}.

Additionally, we ensured that any work discussing XAI without addressing accessibility was excluded from our study. For example, ``XAI for intrusion detection system: comparing explanations based on global and local scope" \cite{hariharan2023} and ``Explainable Artificial Intelligence (XAI) User Interface Design for Solving a Rubik's Cube" \cite{Bradley2022} were not included in our analysis. Table \ref{tab:articles_per_source} displays the percentage of articles initially selected from each database and the number of accepted and rejected papers from each source.

\begin{table}[htbp]
\caption{Papers percentage by source}
\centering
\begin{tabular}{|c|c|c|c|}
\hline
\rowcolor[HTML]{C0C0C0} 
 Sources & Papers Percentage & Accepted Papers (No.) & Rejected Papers (No.)\\ \hline
 Scopus & 4.7~\% & 1 & 33\\ \hline
 ISI Web of Science & 63.3~\% & 0 & 460\\ \hline
 IEEE Digital Library & 3~\% & 0 & 22 \\ \hline
 Google Scholar & 1.2~\% & 0 & 9\\ \hline
 El Compendex & 25.2~\% & 2 & 181 \\ \hline
 ACM Digital Library & 2.6~\% & 1 & 18\\ \hline
 CINAHL & 0.3~\% & 0 & 2\\ \hline
 PubMed & 1.8~\% & 0 & 13\\ \hline
\end{tabular}
\label{tab:articles_per_source}
\end{table}

\subsection{Key Papers and Their Contributions}
As a result of our research, we have selected four papers: two from 2020 and two from 2023. The following text provides a summary of these four papers on AXAI Experiences. 

One of the papers, titled ``Designing Accessible, Explainable AI (XAI) Experiences" \cite{WolfChristine}, explores the intersection of accessibility and XAI. It emphasizes the significance of making complex AI and ML models understandable and accessible, focusing on two main areas: accessibility at the interface and tailoring explanations to diverse and changing user needs. This article uses aging-in-place and mental health support as case studies to illustrate the challenges of making XAI more accessible, particularly for those with specific needs.

The paper ``Accessible Cultural Heritage through Explainable Artificial Intelligence" \cite{Díaz-Rodríguez2020317} aims to improve accessibility in cultural heritage using XAI. By incorporating XAI techniques, the goal is to make art and cultural experiences more inclusive, especially for marginalized communities. The paper discusses the challenges in making cultural heritage accessible to everyone, including people with disabilities, and suggests using XAI to bridge the gap. The proposed methodology uses generative and multimodal models, NLP, and image captioning to create more informative and engaging art experiences. The primary conclusion is that XAI can play a significant role in making cultural heritage more accessible and engaging for diverse audiences, including those who are visually impaired.

The paper titled ``Increasing Transparency to Design Inclusive Conversational Agents (CAs)" \cite{MottaIsabela} explores how AI-based CAs can be used to include marginalized and vulnerable populations such as people with mobility, visual or hearing disabilities, older adults, and those with mental illnesses. The paper highlights the importance of increasing transparency in CAs to be more inclusive and address misperceptions. It also emphasizes challenges such as personalizing transparency levels and creating human-centred knowledge on transparency and explainability. Ultimately, the article concludes that enhancing transparency in CAs can improve learning and performance, address privacy concerns, and promote inclusion.

The article ``A Transparent CAPTCHAS Verification System for Cloud-Based Smart and Secure Applications" \cite{ShahShahUllah} introduces a novel CAPTCHA verification system designed for cloud-based applications. The system employs advanced deep-learning techniques to convert image CAPTCHAS into text, utilizing Explainable AI (XAI) for transparency in the conversion process. This approach enhances accessibility for individuals with disabilities by providing an inclusive verification method online. The system shows promise in CAPTCHA-based authentication, improving security and user experience, especially for those with accessibility needs.

In order to gain insights from the selected research papers, we established correlations among them, as described in \cite{Papaioannou2016systematic}. By doing so, we aimed to identify potential trends and patterns in accessible explainable artificial intelligence (XAI). This approach helps us to understand which topics are gaining popularity and where gaps in the research might exist. Correlating articles is a crucial step in systematic reviews and meta-analyses, which enable us to synthesize the results of various studies to form comprehensive conclusions on a given topic, as explained in Booth et al. \cite{Papaioannou2016systematic}. Additionally, considering the specific field of XAI, the work \cite{Gunning_Aha_2019} provides valuable insights into the current state and future directions of explainable AI, underscoring the relevance of identifying patterns and gaps in this rapidly evolving area.

\subsection{A Comprehensive Analysis of AXAI Research}
Analyzing the selected papers in Table \ref{table:articles_summary}, we identified common themes related to accessibility, inclusion, and explainable artificial intelligence (XAI). However, each paper presented specific focuses and contributions.

\begin{table*}[!ht]
\caption{First summary of selected papers}
\centering
\footnotesize
\begin{tabular}{|L{1.3cm}|L{1.9cm}|L{1.8cm}|L{1.5cm}|L{1.6cm}|L{1.5cm}|L{1.7cm}|L{1.6cm}|}
\hline
\rowcolor[HTML]{C0C0C0} 
\textbf{Authors} & \textbf{Methodological Approach} & \textbf{Technologies or Tools Used} & \textbf{Proposed Design} & \textbf{Target Audience} & \textbf{Paper Context} & \textbf{Paper Contribution} & \textbf{Core Themes} \\
\hline
Wolf et al. \cite{WolfChristine} & User-Centric & Visual explanations and dialogue systems & Accessible and customizable design & Customizable design	People with accessibility needs (blind, low-viz, aging, mental healthcare) & XAI interface accessibility & Discussions, reflections & XAI, accessibility \\
\hline
Diaz-Rodriguez et al. \cite{Díaz-Rodríguez2020317} & User-Centric & Generative, multimodal AI & Adaptive and inclusive design & Physical, cognitive disabilities (non-technical, blind, deaf, elderly) & Cultural heritage (Art) & Opportunities, challenges & XAI, accessibility \\
\hline
Motta and Quaresma \cite{MottaIsabela} & User-Centric & AI Voice Assistants & Transparent and inclusive design & Disabilities (visual, motor, elderly, marginalized) & AI-based Conversational Agents & Discussions, reflections & Transparency, inclusion \\
\hline
Shah et al. \cite{ShahShahUllah} & CNN, User-Centric & Deep-learning techniques, XAI & Accessible and user-friendly design & Disabilities (general) & Image CAPTCHA transformation & System development & Transparency, inclusion \\
\hline
\end{tabular}
\label{table:articles_summary}
\end{table*}

The choice of variables for analysis among the papers in Table \ref{table:articles_summary} is justified by their range and depth in understanding the studies. ``Methodological Approach", ``Technologies or Tools Used", and ``Proposed Design" highlight how research is conducted. They reflect on the validity and applicability of the results. As discussed in \textit{Practice of Social Research} \cite{babbiepractice}, design research and methodologies are crucial for obtaining reliable data. The relevance and impact of studies in diverse settings and specific groups can be determined by considering the ``Target Audience" and ``Paper Context". This is an important point emphasized by \cite{YSYED202363}, who stresses the significance of including diversity, accessibility, and the context in the search. Additionally, the ``Paper Contribution" and ``Core Themes" clarify the research's innovation, added value, and central topics. These factors are fundamental in evaluating the contribution of each article to the field of study.

The papers in Table \ref{table:articles_summary} are strictly correlated in several aspects. The user-centred methodology is a cornerstone in all four articles, with additional convolutional neural network (CNN) techniques in the \cite{ShahShahUllah} study. The user-centric approach prioritizes the needs, experiences and feedback of end users. This methodology is fundamental to developing solutions that are not only technically viable but also practical, useful and accessible to the intended users.

The papers use a significant variety of technologies and tools, such as generative and multimodal AI, AI-powered voice assistants, and deep learning techniques for transparent CAPTCHAS. Two papers (\cite{WolfChristine} and \cite{ShahShahUllah}) are directly correlated regarding the use of explainable techniques, while two others (\cite{WolfChristine} and \cite{MottaIsabela}) are related due to the use of conversational technologies. This diversity of technology reflects the potential of various AI approaches to improve accessibility and inclusion and the specific challenges they seek to solve within their application context.

All the papers on the proposed design aim to address the challenge of making technology more accessible and inclusive. Designing systems and interfaces that cater to a wide range of users, including those with different needs, is a significant concern. The focus is on making AI-based systems transparent and accessible, which emphasizes the importance of considering how these technologies can be adapted to meet the needs of all users.

All papers in Table \ref{table:articles_summary} have as target audience individuals with different accessibility needs. Considering that this research is focused on the study of accessible XAI for those with sight loss, we have the paper \cite{WolfChristine} covering blind and low-visibility (low-viz) users; paper \cite{Díaz-Rodríguez2020317} including blind or visually impaired users; for \cite{MottaIsabela}, people with visual disability; and paper \cite{ShahShahUllah} does not explicitly mention individuals with sight loss. However, the final system could be valuable to the community despite lacking details on the use of XAI for decisions to transform the CAPTCHA image into text, which would help us understand if the approach could also work for blind and low-vision individuals.

Regarding the paper context, each article applies accessibility and XAI concepts in different domains: AI interfaces (\cite{WolfChristine}), cultural heritage (\cite{Díaz-Rodríguez2020317}), conversational agents (\cite{MottaIsabela}) and CAPTCHAS verification systems (\cite{ShahShahUllah}). This diversity of applications shows the scope and relevance of accessibility and inclusion issues in various areas of technology and how different strategies and technologies can be employed to address these challenges in specific contexts.

The analyzed papers cover a wide range of topics, from discussing AI's opportunities and challenges to developing specific systems. These contributions have significantly improved the field of AI and accessibility, providing valuable insights into how technology can be designed and implemented in a more inclusive and accessible way. It is worth noting that while the studies emphasize the need for an accessible XAI, they do not necessarily propose any concrete solutions based on the discussions. Only one paper, \cite{ShahShahUllah}, presents the implementation of a transparent system based on an accessible XAI. However, it is still unclear how AI explanations are incorporated into this system, as well as the tests with end-users.

As shown in the last column of Table \ref{table:articles_summary}, the core theme of all the works analyzed rotates around the same topics: XAI or transparency and accessibility or inclusion. This confirms how related the studies are to this research.

\begin{table}
\caption{Second summary of selected papers}
\centering
\footnotesize
\begin{tabular}{|L{1.3cm}|L{2.6cm}|L{2.5cm}|L{1.8cm}|L{2.2cm}|}
\hline
\rowcolor[HTML]{C0C0C0} 
\textbf{Authors} & \textbf{Problem Addressed} & \textbf{Solution Proposed} & \textbf{Method Used} & \textbf{Identified Gaps} \\ \hline
Wolf et al. \cite{WolfChristine}& AI/ML complexity & Accessible XAI interfaces, explanations & Discussion, case studies & Adaptable explanations needed \\ \hline
Diaz-Rodriguez et al. \cite{Díaz-Rodríguez2020317} & Cultural heritage inaccessibility & XAI for accessible, engaging art & Generative models, NLP, captioning & Integration with cultural databases \\ \hline
Motta and Quaresma \cite{MottaIsabela} & AI-based CAs exclusivity & Transparent, inclusive CAs & User interaction analysis & Transparency settings customization \\ \hline
Shah et al. \cite{ShahShahUllah} & CAPTCHA inaccessibility & Transparent CAPTCHA via deep learning & Deep learning, XAI techniques & Broader web accessibility integration \\ \hline
\end{tabular}
\label{table:articles_summary2}
\end{table}

The problems and solutions presented in Table \ref{table:articles_summary2} vary significantly. The first paper \cite{WolfChristine} focuses on making XAI understandable and accessible, discussing interface issues and customizing explanations for different users. The second \cite{Díaz-Rodríguez2020317} expands accessibility to cultural heritage using XAI, proposing techniques with generative models and image descriptions. The third \cite{MottaIsabela} highlights the importance of transparency in conversational agents for inclusion, suggesting personalization of transparency. The fourth \cite{ShahShahUllah} introduces a transparent CAPTCHA verification system, using deep learning to transform image CAPTCHAs into text, improving accessibility.

The papers analyzed highlight the growing importance of accessibility and explainability or transparency in AI, focusing on diverse applications and contexts. One trend is the search for personalized and inclusive solutions that meet different user needs. A noted gap is the need for more research and effort into how AI explanations can be adapted for users with different types of accessibility and deeper integration of it with existing assistive technologies.

\subsection{XAI Taxonomic Analysis/Relationships - Accessibility and Disability }

It has been established above that XAI discussions are lacking the accessibility and disability perspective, but where it is possible, our analyses would include human centered aspects that might enhance accessibility for persons with visual disability.  In this section, we aim to relate our findings to XAI taxonomies published in reputable venues such as Elsevier Science Direct, Web of Science and IEEE Xplore Digital Library. Several works on XAI taxonomy were identified and collected and used for this analysis. The exclusion criteria for this search include: conference papers, articles prior to 2020, articles that did not include taxonomies or extensive classifications of XAI. Conference papers were excluded because we only want to utilize sources that provided sufficient depth in the discourse of XAI taxonomies and classifications. To further describe depth here, we mean the articles that reviewed and classified both XAI algorithms, technologies/systems and assessment approaches for users. Table \ref{table:Taxonomic_analysis1} contains the results of the first group of findings in this section. 

\begin{table*}
\caption{Taxonomic analysis for accessibility, trust and transparency}
\footnotesize
\begin{tabular}{|L{1.2cm}|L{1.5cm}|L{1.5cm}|L{10.5cm}|}
\hline
\rowcolor[HTML]{C0C0C0}
\textbf{Authors} & \textbf{AI and XAI \newline Method} & \textbf{Assessment Approach} & \textbf{XAI Taxonomies (sections where XAI methods were discussed)} \\ \hline
Wolf et al. \cite{WolfChristine} & Visual explanations (VE); DNN & None & -PHE (Joint prediction and explanation(JPE))-VE\cite{ALI2023101805}; \newline -Perturbation, BackPropagation-VE\cite{Das2020}\newline -PHE-VE(Arrieta \cite{Arrieta2020}); \newline -Feature influence (relevance) methods, Gradient-based Explanations, Explanation presentation-VE\cite{Eldrandaly2023}; \newline -Post-Hoc Interpretability and Explanations-VE\cite{Kamath2021}; \newline -PHE techniques-VE \cite{Nazar2021} \newline -PHE, Model-Agnostic-VE \cite{Belle2021} \newline -Presentation Format-VE \cite{Martins2023} \\ \hline
Diaz-Rodrigues et al. \cite{Díaz-Rodríguez2020317} & VQA, Image captioning (IC), GAN & None & -PHE (JPE)-VQA \cite{ALI2023101805}; \newline -Perturbation-VQA \cite{Das2020} \newline -Dialogue approaches-VQA \cite{Eldrandaly2023} \newline -Feature Importance, Provenance-Based-VQA \cite{Kamath2021} \newline -Explainability in deep learning-IC \cite{Arrieta2020} \newline -Tools and Libraries, Intrinsic (Attention)-IC \cite{Kamath2021} \newline -ML Characteristics, Accountable; Explainability techniques-IC\cite{Nazar2021}  \\ \hline
Motta and Quaresma \cite{MottaIsabela} & AI Agents & None & -Cognitive psychological measures; PHE (Game theory methods)\cite{ALI2023101805} \newline -Explanations for AI security, XAI and adversarial ML \cite{Arrieta2020} \newline -Dialogue approaches; Explanation presentation; XAI Evaluation measures \cite{Eldrandaly2023} \newline -Reinforcement Learning \cite{Nazar2021} \\ \hline
Shah et al. \cite{ShahShahUllah} & LSTMRNN and CNN & None & -Memory networks; Model explainability-LSTM \cite{ALI2023101805} \newline -Local Explanations (SHAP)-LSTM \cite{Das2020} \newline -DL Explainability, Recurrent Neural Networks-LSTM \cite{Arrieta2020} \newline -Experimental evaluation of XAI methods-LSTM \newline -Intrinsic (Attention)-LSTM \cite{Eldrandaly2023} \newline -ML/DL Accountability-LSTM \cite{Nazar2021} \newline \newline -Explainability through architectural adjustments (EAA), JPE, Explainability through regularization, PHE (Class Activation Map, Gradient-weighted CAM), Vanilla Gradient; Visualization methods-CNN \cite{ALI2023101805} \newline -Perturbation-Based (DeConvolution nets for Convolution Visualizations, BackPropagation- or Gradient-Based (Saliency Maps, GradientCAM)); Local Explanations (Activation Maximization, Layer-wise Relevance BackPropagation); Global Explanations (Class Model Visualization, Neural Additive Models)-CNN \cite{Das2020} \newline Explainability in DL (Convolutional Neural Networks), PHE (Feature relevance)-CNN \cite{Arrieta2020} \newline
-Computational measures, Experimental evaluation of XAI methods - CNN\cite{Eldrandaly2023} \newline
-Intrinsic: Attention-based neural networks (ABNN) (Perturbation), Gradient/Backpropagation (Activation Maximization, Class Model Visualization, Saliency Maps) - CNN \cite{Kamath2021} \newline
-Convolutional Neural Network; Explanation by simplification - CNN \cite{Nazar2021} \\ \hline
\end{tabular}
\label{table:Taxonomic_analysis1}
\end{table*}

To conduct the analysis for the taxonomic relationships of the above findings on accessibility and XAI, we commence with the paper by Wolf et al. \cite{WolfChristine}, which is actually a newsletter and not research/technology paper, so eliciting actual XAI technologies required some investigation. The reason behind this investigation is that it used case studies and reflections, XAI and AI/ML was described in broad terms and later discussed in line with the age-in-place and mental health administration. However, the article based its assertions on \textbf{\textit{visual explanations (VE)}} while citing the work by Hendricks et al. \cite{Hendricks2016}. So, an in-depth analysis of this work provided us the necessary information to understand the classification of the AI-based approach used. Ali et al. \cite{ALI2023101805} mentioned VEs under section 7.3, i.e., “\textit{Joint prediction and explanation}”, for dataset training while comparing Park et al., \cite{Park2018}’s model of explainability to the Teaching Explanations for Decisions (TED) framework. Deep Neural Networks (DNN) was employed by Hendricks et al. \cite{Hendricks2016} for VEs in the context of recognition and detection of objects. Specifically, this method demands explanations in a textual format alongside the class labels during training to be able to provide distinct class-dependent VEs of the inputted image predictions during testing. 

Additionally, the possibilities of utilizing VEs were again discussed under Randomized Input Sampling to Provide Explanations (RISE) and Gradient-weighted CAM (Grad-CAM) within Ali et al \cite{ALI2023101805}’s proposed taxonomy for post-hoc explainability (PHE).  It is noteworthy that this form of explainability is organized around six key attributes and methods namely: example-based explanation, visualization, knowledge extraction, attribution, game theory and neural methods. RISE and CAM are classified under attribution methods. In Arrieta et al. \cite{Arrieta2020}, VE's was also discussed under post-hoc explainability.  In addition, DNNs was discussed under several sections which include Hybrid explainable models, Explainability through regularization, RISE, Back propagation methods and Model distillation, etc. The extensive usage of DNNs is premised on the fact that it is very useful when obtaining meaningful information from very complicated datasets and that networks with considerable depths facilitates improved decision-making compared to shallow ones. Since the article is a newsletter of accessible XAI reflections without usability assessments, it was not possible to discuss any of the mentioned XAI assessment approaches in Ali et al. \cite{ALI2023101805}. 

The paper by Dıaz-Rodrıguez and Pisoni~\cite{Díaz-Rodríguez2020317} posed issues, important research questions and hypothesis for art accessibility using XAI approaches. Here, we present taxonomic analysis of identified XAI methods alongside the relevant papers of XAI taxonomy and classification. Hypothesis 1 is about providing assistance through AI-based models that can generate content- and interpretation-wise explanations. The consideration for viable solution is \textit{\textbf{image captioning}}. Ali et al. \cite{ALI2023101805} did not include this concept in their taxonomy. However, captioning (image, video) and its techniques were discussed and implemented in Stangl et al. \cite{Stangl2023}. While hypothesis 2 asked whether XAI can explain art, research question 2 asked if XAI can both produce and distinguish between explanations of content and form. The consideration for viable solution is \textbf{\textit{visual question answering (VQA)}}. VQA in ref \cite{ALI2023101805} was briefly mentioned under section 7.3 i.e., “Joint prediction and explanation”. Research question 3 asked if art explanations can be produced on request through VQA – this concept is also on-demand interaction implemented in Stangl et al. \cite{Stangl2023} using AD systems.  The last method identified in the paper is \textit{\textbf{Generative adversarial networks (GANs)}} and multimodal CNN presented under the section for explaining visual art. While the concept of CNN was widely seen in Ali et al. \cite{ALI2023101805}, we did not find GAN and multimodal CNN. The article is a review paper for art accessibility using XAI and it lacked discussion of assessments, therefore was not included here.

The paper by Motta and Quaresma \cite{MottaIsabela} involved AI-based conversational agents and voice assistants. Intelligent agents that aid conversations are not widely mentioned in Ali et al. \cite{ALI2023101805}, perhaps due to their focus on technical aspects of XAI developments rather than human-centered needs and requirements. The only relevant mention of \textbf{\textit{AI agents }}was under assessment of explanations and specifically, under “Cognitive psychological measures”. Here, the theme is user understanding of AI-based agents and algorithms. Further investigations of references revealed the study by Penney et al. \cite{Penney2018} titled “Toward foraging for understanding of StarCraft agents”. This paper is game-based as participants played the real-time strategy game and it is noteworthy that post-hoc explainability of ref \cite{ALI2023101805} included the game-theory methods, which are Shapley Values and Shapley Additive Explanation. The rationale behind the lack of more information on agents within Ali et al. \cite{ALI2023101805} is because their interest was unfortunately not accessibility and inclusion. Generally, from our discussions it may seem that the concepts of voice assistants and audio descriptions are same because of sound. Truly, there are significant similarities which are accessibility (improvements for persons who are visually impaired through additional interaction modes and conveying the requisite information); NLP (understand questions or user commands for voice assistants or produce the right ADs) and involves AI/ML algorithms. Both technologies differ in terms of functionality and focus of the content, for example, Siri/Alexa can be used for lots of purposes, but ADs are for describing visual content.

The article by Shah et al. \cite{ShahShahUllah} focused on XAI and enhancing transparency and inclusion for persons with disability by the implementation of a Completely Automated Public Turing Test to Tell Computers and Humans Apart (CAPTCHAS) Verification System. However, a Connectionist Temporal Classification (CTC) layer was incorporated into their system. This makes it easier to learn from sequence to sequence, enabling direct predictions of the CAPTCHA text and eliminating the need for intricate alignment. Shah et al. \cite{ShahShahUllah} utilized the CTC approach, and the aim is to create the textual version of the CAPTCHA while training by using the text labels and the knowledge that has been learnt. The proposed system applied AI methods such as LSTMRNN and CNN, which were also identified in Ali et al. \cite{ALI2023101805} taxonomy but the CTC layer, which implements XAI was absent. It is noteworthy that text generation may be somewhat similar to functions which include; image caption generation, text summarization and text on screen \cite{Stangl2023}. Text production and processing are involved in all activities, whether they are producing descriptions in plain language, compressing text, and displaying information graphically. Each of them make use of natural language processing approaches, and some of the tasks may require the use of ML algorithms.

\section{Promising Avenues for Exploring AXAI: Accessible Human-Centered Computing Techniques for Explainability}

In this section, we provide a summary of promising avenues for exploring human-centered computing techniques for explainability. This is because, we believe these avenues hold potential to impact the lives of individuals who are blind and partially-sighted. 

First, we refer to accessible human-centred computing techniques for explainability, wherein findings include conference papers published in the ACM SIGACCESS Conference on Computers and Accessibility and others. Generally, these types of systems have progressed considerably, merging knowledge representation and reasoning, natural language processing, computer vision, and XAI. The findings are as follows: \cite{Stangl2023}, \cite{Ihorn2021}, \cite{Bodi2021}, \cite{Yuksel2020}, \cite{Yuksel2020human}. Note that these individuals who were referred to as blind and low vision (BLV). 

Second, the lives of people with visual impairments were seemingly improved by audio description (AD) systems, according to research on accessibility and HCC \cite{Stangl2023}. However, within the framework of this research, this emphasizes how important it is to comprehend and investigate audio XAI technological advances. The pertinent audio XAI techniques that may be applied to the development of audio explanations were enumerated by Akman et al. ~\cite{Akman2024}. These methods include: Layer-Wise Relevance Propagation (LRP), Discrete Fourier Transform-Layer-Wise Relevance Propagation (DFT-LRP); Cough-Local Interpretable Model-agnostic Explanations (Cough-LIME); loudness/ Non- negative Matrix Factorization (NMF) decomposition, Spleeter, Surrogate self-explainable model NMF decomposition; and the combination of LIME, Causal, Statistical Fault Localisation (SFL).

The analysis of these techniques using the format above (findings and taxonomic analysis) will be explored in subsequent research. However, it seems clear that these techniques present promising avenues to solve complex issues of accessibility for individuals who are blind and partially-sighted.

\section{Conclusion, Recommendations and Research Road-map}
The motivation for this review was drawn from the documented issues and barriers faced by individuals with visual disabilities \cite{Bell2018}, \cite{Crudden1999}, \cite{Kirchner2005}, \cite{Mcdonnall2019}, \cite{Mcmillen2017}, \cite{Shaw2007}. These barriers are particularly evident in technologically advanced settings where essential information is communicated visually, resulting in social, educational, and career marginalization and perhaps affecting their independence, autonomy, and well-being \cite{Stangl2023}. 

The current state of XAI research and accessibility can be compared to outdated approaches in designing and developing applications, web pages, and digital tools, with no accessibility guidelines for developers to follow. In recent times, the importance of adhering to guidelines cannot be overstated \cite{Filipe2023}, \cite{Ara2023}. Therefore, we advocate for the creation of accessibility guidelines for the development of AI-based systems. 

\subsection{New research directions for XAI}
Considering our findings and their implications, we now provide ideas and important research directions for enhancing the accessibility of XAI for the disability community.
\begin{enumerate}
    \item We emphasized the importance of involving the disability community in XAI discourse. However, our paper focused mainly on visual disabilities such as blindness and low vision, suggesting the need for further research into other types of disabilities. For some mental and sensory disabilities, we can only imagine the complexities it might entail considering the complex XAI landscape.

    \item With further research and implementation, if transitioning from audio descriptions to audio explanations proves to be practical and effective, there may be a necessity to consider the processes involved in authoring and evaluating audio explanations (AE). However, we need to understand the meaning of ``minimum" in MVD \cite{Stangl2023} and how it improves access for the disability community, with explanations that would positively impact them. This will depend on extended research confirming the significance of the phrase ``\textit{\textbf{minimum viable explanations}}" to disability and accessibility. It refers to understanding the least amount of explanation that produces the best results, as it has been found that too much information can cause cognitive overload, reducing a user's trust and understanding of the system\cite{Schmidt2020, Hudon2021}.  

    \item In addition to the subjective nature of explanation, Pisoni et al. \cite{Pisoni2021} aligned with advocating for more thoughts towards interpretability \cite{Hanif2023} and the expertise of the audience \cite{Severes2023} in question. Since individuals with visual disability constitute the population of interest here, there is a need to consider literacy level while conceiving and implementing explanations. Based on the papers reviewed, potential future research could concentrate on creating multisensory interfaces that can explain AI decisions and tailor AI explanations to better cater to users' needs with a broader range of disabilities and preferences. It is worth exploring how AI explanations can be more contextual and situational, adapting dynamically to the user's context. Furthermore, exploring interactive approaches, where users can request additional or more detailed clarification about AI decisions, could enhance comprehension and trust in the technology.

    \item The role of human-in-the-loop (HITL) \cite{Yuksel2020} approach for explanations, XAI, accessibility and disability needs to be investigated, since Stangl et al. \cite{Stangl2023} had successes using it for verification and revision of AI-generated captions and descriptions.  While Stangl et al. \cite{Stangl2023} confirmed that the HITL approach has made AI-based AD more acceptable \cite{Morash2015}, they added that the process nonetheless demands infrastructure, funding, and time to operate at scale and may offer variable results.

    \item In our search to discover approaches to assist persons with visual disabilities, a system such as FitVid came to the fore. FitVid \cite{Kim2022} is a platform that facilitates video contents that are both customizable and responsive. In the context herein, extended research is required to evaluate if XAI models can be made accessible using FitVid. They discussed “advanced accessibility with content customization” and the basic idea involves enhancing accessibility for several circumstances and conditions. There were specific mentions of disability populations such as low vision, colour blindness and dyslexia as candidates of content adaptation. We believe that this is an attempt at inclusivity, which proposes the addition of color palette settings for people with color blindness rather than a just dark theme. 

    \item Inclusion of users who are living with visual disabilities is very important for the XAI/accessibility research. Several approaches for assessing XAI-based systems were provided in Ali et al. \cite{ALI2023101805}, and we believe they may positively impact this area of research depending on the available funding, time and infrastructure. As a matter of fact, studying the effectiveness of different explanations and how users interact with them in real-world scenarios could provide valuable insights. Involving users with disabilities in the XAI design process through co-creation can also lead to more inclusive and practical solutions.
\end{enumerate}

For individuals who are blind, deaf blind, or with other forms of vision loss, it is essential to develop techniques that utilize sound narratives or auditory descriptions \cite{AlexaSiu2022} to clarify AI's functioning and decisions. Additionally, it is crucial to ensure that AI explanations are compatible and can be integrated with assistive technologies currently used, such as screen readers. Making AI systems' explanations and operations understandable by people with visual disabilities increases their autonomy, allowing them to make informed decisions based on a clear understanding of how technologies work and impact their lives.

Upon analyzing the papers in the previous sections, it was found that there has been a lack of attention given to providing explanations of AI for people with sight loss. To address this issue, the notion of Accessible Explainable Artificial Intelligence, AXAI, has been proposed. Achieving AXAI will ensure that AI systems are easily understandable and accessible to everyone, thus promoting digital inclusion. It is crucial to advance research and development in this area because it is not just a matter of accessibility but also of equity and inclusion. We must commit to research, development, and collaborative testing with the sight loss community to create solutions that meet their specific needs and preferences.

\section*{Acknowledgements}
This project has been made possible by Accessibility Standards Canada / the Government of Canada. The research was also undertaken, in part, thanks to funding from the Canada Research Chairs Program.

%Bibliography
\bibliographystyle{unsrt}  
\bibliography{main}

\end{document}